\documentclass{article}

\usepackage{arxiv}
\usepackage[utf8]{inputenc}
\usepackage[version=4]{mhchem}
\usepackage{caption}
\usepackage{subcaption}
\usepackage{tabularx}
\usepackage{array}
\usepackage{booktabs}
\usepackage{upgreek}

\usepackage[T1]{fontenc}    
\usepackage{hyperref}       
\usepackage{url}            
\usepackage{cleveref}       
\usepackage{graphicx}
\usepackage[square,numbers,sort&compress,comma]{natbib}
\usepackage{indentfirst}
\usepackage{multirow}
\usepackage{siunitx}
\begin{document}

\title{Comparing Forward and Inverse Design Paradigms: A Case Study on Refractory High-Entropy Alloys}
\author{Arindam Debnath\\
Department of Materials Science and Engineering\\
Pennsylvania State University\\
University Park, PA 16802\\
\And
Lavanya Raman\\
Department of Materials Science and Engineering\\
Pennsylvania State University\\
University Park, PA 16802\\
\And
Wenjie Li\\
Department of Materials Science and Engineering\\
Pennsylvania State University\\
University Park, PA 16802\\
\And
Adam M. Krajewski\\
Department of Materials Science and Engineering\\
Pennsylvania State University\\
University Park, PA 16802\\
\And
Marcia Ahn\\
Department of Materials Science and Engineering\\
Pennsylvania State University\\
University Park, PA 16802\\
\And
Shuang Lin\\
Department of Materials Science and Engineering\\
Pennsylvania State University\\
University Park, PA 16802\\
\And
Shunli Shang\\
Department of Materials Science and Engineering\\
Pennsylvania State University\\
University Park, PA 16802\\
\And
Allison M. Beese\\
Department of Materials Science and Engineering\\
Department of Mechanical Engineering\\
Pennsylvania State University\\
University Park, PA 16802\\
\And
Zi-Kui Liu\\
Department of Materials Science and Engineering\\
Pennsylvania State University\\
University Park, PA 16802\\
\And
Wesley F.~Reinhart\\
Department of Materials Science and Engineering\\
Institute for Computational and Data Sciences\\
Pennsylvania State University\\
University Park, PA 16802\\
\texttt{reinhart@psu.edu}\\
}
\date{\today}

\maketitle


\begin{abstract}
    
The rapid design of advanced materials is a topic of great scientific interest.
The conventional, ``forward'' paradigm of materials design involves evaluating multiple candidates to determine the best candidate that matches the target properties.
However, recent advances in the field of deep learning have given rise to the possibility of an ``inverse'' design paradigm for advanced materials, wherein a model provided with the target properties is able to find the best candidate.
Being a relatively new concept, there remains a need to systematically evaluate how these two paradigms perform in practical applications.
Therefore, the objective of this study is to directly, quantitatively compare the forward and inverse design modeling paradigms. 
We do so by considering two case studies of refractory high-entropy alloy design with different objectives and constraints and comparing the inverse design method to other forward schemes like localized forward search, high throughput screening, and multi objective optimization.

\end{abstract}



\section{Introduction}\label{into}

The discovery of advanced materials that can meet the requirements of the current and future generations for applications like energy generation and storage, water purification, and carbon sequestration is a subject of significant scientific interest.
However, the design of new materials with desired value of properties is a challenging task due to the large and mostly uncharted design space and the difficulty in predicting the non-linear relationships between structure, property, and processing parameters \cite{Debnath2021}.
While the three traditional paradigms of materials design -- Edisonian trial and error, theoretical models governed by physical and chemical rules, and computer simulations -- have been successful over the past decades, implementing these strategies for modern materials design problems has become increasingly challenging for several reasons.  
The trial-and-error method of materials design relies heavily on serendipitous discovery and often take 10–20 years to develop a new material with desired properties, from initial research to its deployment in an application \cite{liu2017materials}. 
The complexity of the theoretical models means that analytical solutions are oftentimes difficult to derive \cite{agrawal2016perspective}.
Simulation methods are more ideal for tasks like prediction of phase formation, phase stability and crystallization kinetics, and it is computationally expensive to use them for optimizing compositions for mechanical properties \cite{liu2018ocean, wen2019machine}.

However, the rich ecosystem of data from these traditional paradigms and the emergence of data-driven approaches like Machine Learning (ML) have unlocked a fourth paradigm of materials discovery. 
This usually involves building forward models that can predict the property of a material given relevant information like the processing, composition, or structure parameters.
As these models take a fraction of the time taken by simulations or experiments, they can help researchers plan future experiments and simulations.
An added benefit of these models is that they can be also used for materials design and discovery. 
There are two ways in which this can be performed. 
The first involves predicting the properties of the compositions in the entire chemical space and then conducting a search to find the most promising materials with the desired properties.
The second involves performing global optimization in the chemical space, wherein the composition of a material system is optimized by trying to maximize or minimize the value of a property, subject to certain constraints.
This requires multiple calls to the forward model, and hence having a fast model is advantageous.
However, while the first method can be thought of as a brute-force approach and would require considerable time and computational resources, the second method could find it challenging to obtain the optimal solution as the property space may be discrete.

\begin{figure}[ht]
     \centering
        \includegraphics[width=\textwidth]{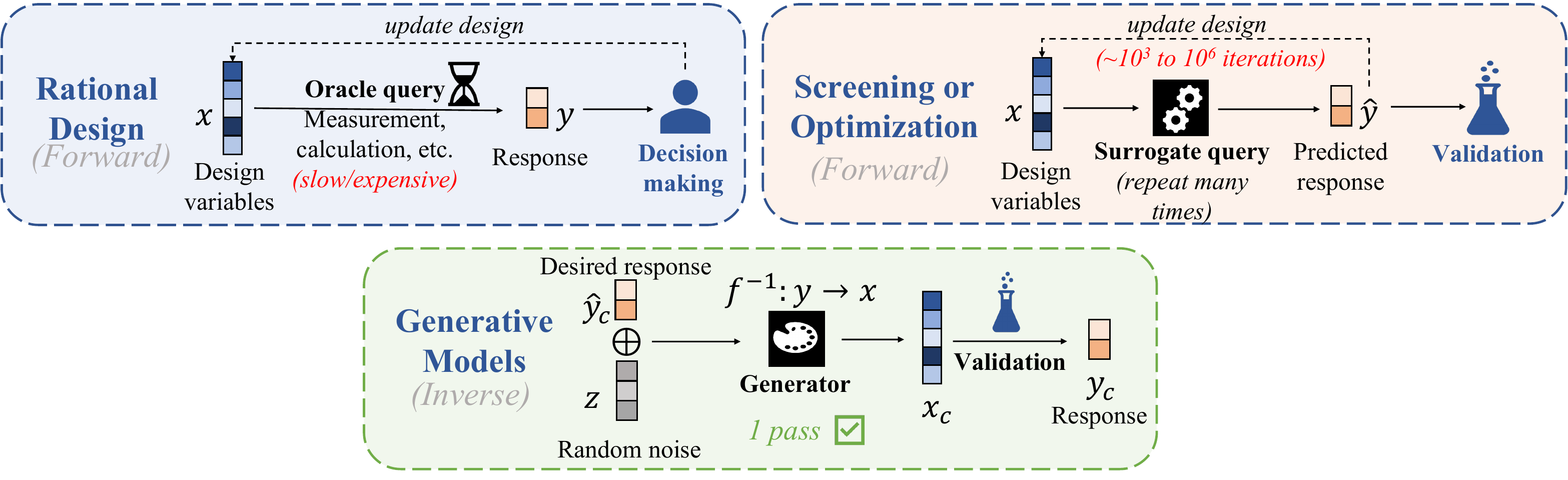}
        \caption{A schematic of the forward and inverse design methods for materials design}
        \label{fig:forward_v_inverse}
\end{figure}

Another promising development in the field of ML are generative models that learns an estimate of the distribution of the data to create new samples that closely resemble the training data. 
Some examples of generative models include Restricted Boltzmann Machines, Gaussian Mixture Models, Deep Belief Networks, Variational Autoencoders (VAE), and Generative Adversarial Networks (GAN) \cite{fischer2012rbm, reynolds2009gmm, hinton2009dbn, kingma2019introduction, goodfellow2016deep}.
They are increasingly being used in the domain of materials science for designing materials like optical meta-materials,  and bulk and thin-film inorganic materials \cite{yeung2020global, dan2020generative, dong2020inverse, Debnath2021}.
Even though they show a lot of promise, there is a need to quantitatively evaluate these methods against the forward model aided materials design process.

Therefore, in this paper, we have compared the forward and the inverse design methods using the case of High-Entropy Alloy (HEA) design as the material system of interest. 
HEAs are a special class of alloys that, unlike conventional alloys, contains multiple principal elements (usually 5 or more) with each having atomic fractions between 5\% and 35\% \cite{Cantor2004,Yeh2004,Senkov2018}.
Interest in them has intensified as they display a combination of exceptional properties like superior strength and room temperature ductility, which is not attainable in conventional alloys \cite{Klimenko2021machine,Jung2021}. 
Particularly, HEAs made using refractory metals have been touted as potential candidates for application in jet engines and gas turbines due to their ability to retain their mechanical properties at high temperatures.
However, only a limited number of the discovered HEAs have properties exceeding those of the current generation of turbine materials, highlighting the importance of designing new HEAs that can meet these requirements.
Unfortunately, designing HEAs with target properties is challenging due to the large and largely unexplored design space, which complicates the prediction of structure-property-processing relationships -- especially at higher temperature regions.
Additionally, the three traditional paradigms of materials design cannot be employed for HEA design without expending significant time, human, and computational resources, prompting researchers to look towards alternative strategies like generative models to accelerate the alloy design process.
In this study, we evaluate these design strategies on their ability to generate refractory HEA candidates satisfying some design parameters using a surrogate model as the ground truth. 
We first considered a case of satisfying a single design criteria by generating novel alloys with a target value of UTS.
We next considered a case of broadening the design criteria by including the price and density of the refractory HEAs as additional targets to satisfy.
For both these cases, we have used a conditional Generative Adversarial Network (cGAN) as our generative model opf choice. 

\section{Results and Discussion}

\subsection{Single condition design case}

We have been able to demonstrate previously that cGANs can be used to guide refractory HEA design and were able to generate candidates with some targeted properties \cite{Debnath2021}.
However, the previous work was a proof-of-concept, where we arbitrarily considered two properties (shear modulus and fracture toughness) as the conditioning provided to the cGAN. 
One of the design specification that we are interested in meeting is that the refractory HEAs should have UTS above 0.4 GPa at $1200^{\circ}$C.
For this scenario, we trained a cGAN with the architecture described in Section~\ref{gen_model} with only the UTS at $1200^{\circ}$C.
The input to the cGAN model is therefore a five dimensional vector (four latent dimensions plus one conditioning value).
The trained generator was then used to identify a composition, 
\ce{Cr20Hf18Mo6Nb6Ti7W36Zr7} (which we will refer to as $C_0$) with UTS at $1200^{\circ}$C equal to 0.43 GPa.
While the generator could be used to generate a composition with a higher UTS value, we intentionally chose this composition to highlight a particular challenge that comes with alloy design -- incorporating feedback from experiments without having to start over.
The decision to choose the $C_0$ was due to the high W content in the composition, which has high melting point and can easily lead to the inhomogeneity in local chemical composition and microstructure. 
This has been observed in another HEA containing such a similarly high W content. 
Such inhomogeneous feature in the synthesized sample is expected to result in poor agreement between the expected and experimental UTS.

As the W content is not a part of the conditioning of the trained cGAN, one might suggest simply including W content as an additional conditioning to the cGAN architecture and training the model from scratch, which would theoretically allow the cGAN to control the W content in the generated samples.
However, while it would be feasible to do so from a mathematical perspective, the cGAN output already includes the atomic fraction of W in the composition, which would create a logical paradox: if elemental composition is provided as input to the model, what then will be the output of the generative model?
Atomic fraction of elements can only be either an input or output to the model, but not both (at least without being reduced to an identify function).
This motivated the development of an alternate strategy to guide the output from the cGAN generator towards desired compositions without changing the generative model.

\begin{figure}[ht]
    \centering
    \includegraphics[width=0.4\textwidth]{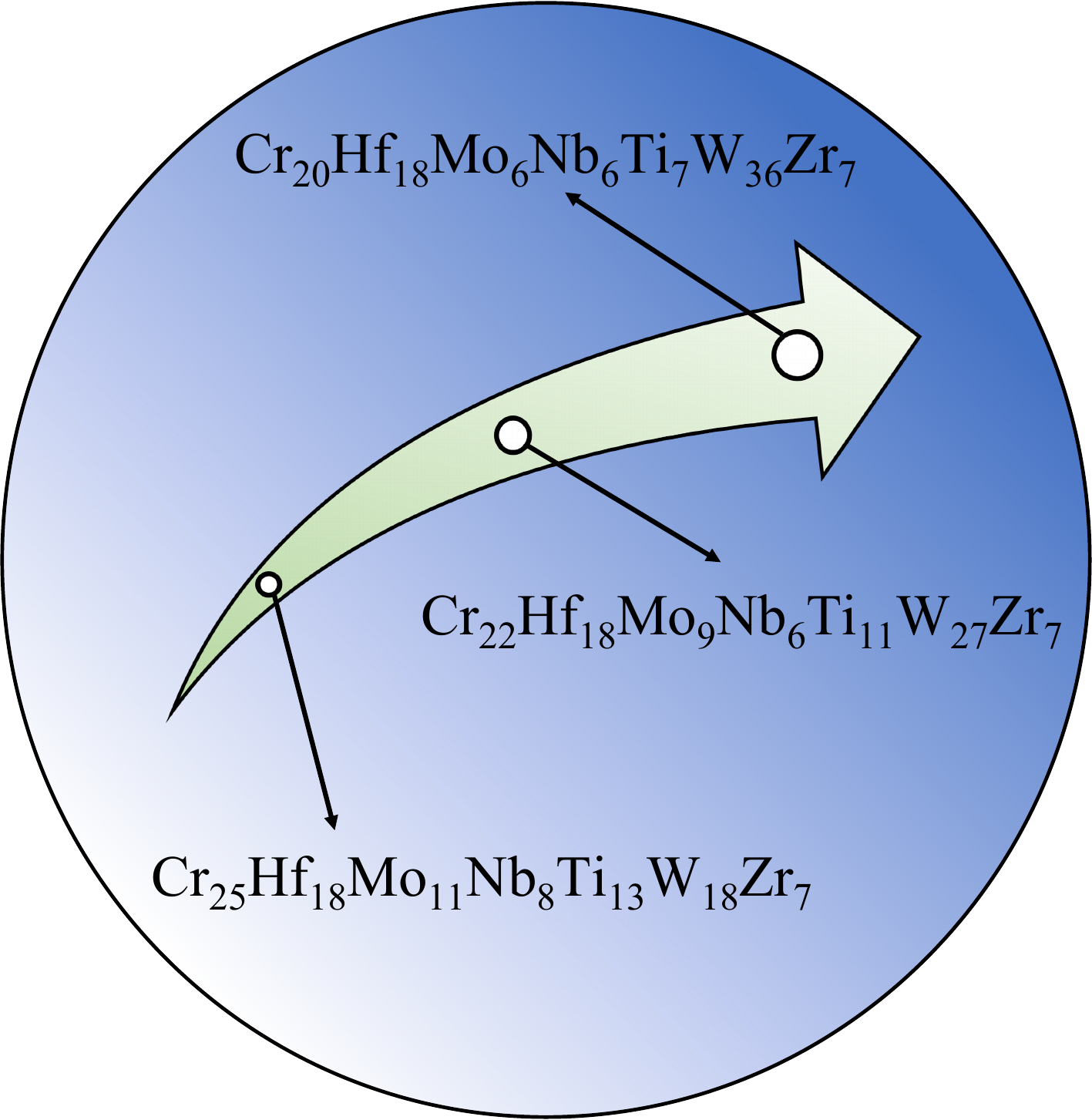}
    \caption{A schematic showing how latent space arithmetic can be used to manipulate the atomic fraction of samples. The circle represents the latent space, with the intensity of blue indicating the W atomic fraction in the composition (light blue regions = low W, dark blue regions = high W). The green arrow shows the direction of maximum variance of W (also called the concept of W, $v_W$), and by traversing along that direction, it is possible to achieve compositions with higher W atomic fraction}
    \label{fig:concept_vector_schematic}
\end{figure}

While the output of latent-variable generative models can be biased by using conditional vectors, another way to control the output is through latent space arithmetic.
As similar observations are grouped together in the latent space, it is possible to identify regions that correspond to a particular state of the sample, as can be seen in Fig.~\ref{fig:concept_vector_schematic}.
If we desire to transform the sample from one state to a different state, we can achieve this by tracing a path between the latent variables corresponding to the two states.
While the path between states need not be linear in principle, vectors are commonly employed in deep learning literature.
By modifying the latent code corresponding to the direction of change from the latent variable of the original sample, we can also generate intermediate samples that smoothly interpolate between the two states \cite{radford2015unsupervised}.

For our particular problem, the first task was finding the direction of maximum variance of W in the latent space, which we will refer to as the concept vector henceforth.
To determine the concept vector, we first decided to check the correlation of W content of the generated compositions with the individual dimensions of the latent codes, $z_i$, used for generating the compositions using the Pearson correlation coefficient $R$.
We did so by considering 10,000 randomly generated compositions from the cGAN.
We observed a maximum $R$ value of 0.55, indicating that there was no strong correlation of W with any of the latent dimensions.
This then prompted us to use a linear regression (LR) model to predict the W content of the generated composition as a function of the latent codes used for generating the 10,000 generated compositions.
The idea behind this is that the coefficient from the LR model would be concept vector for W content.
The implementation of the LR model used for this purpose was from the \texttt{sklearn} Python package \cite{sklearn}. 

\begin{figure}[ht]
    \includegraphics[width=\textwidth]{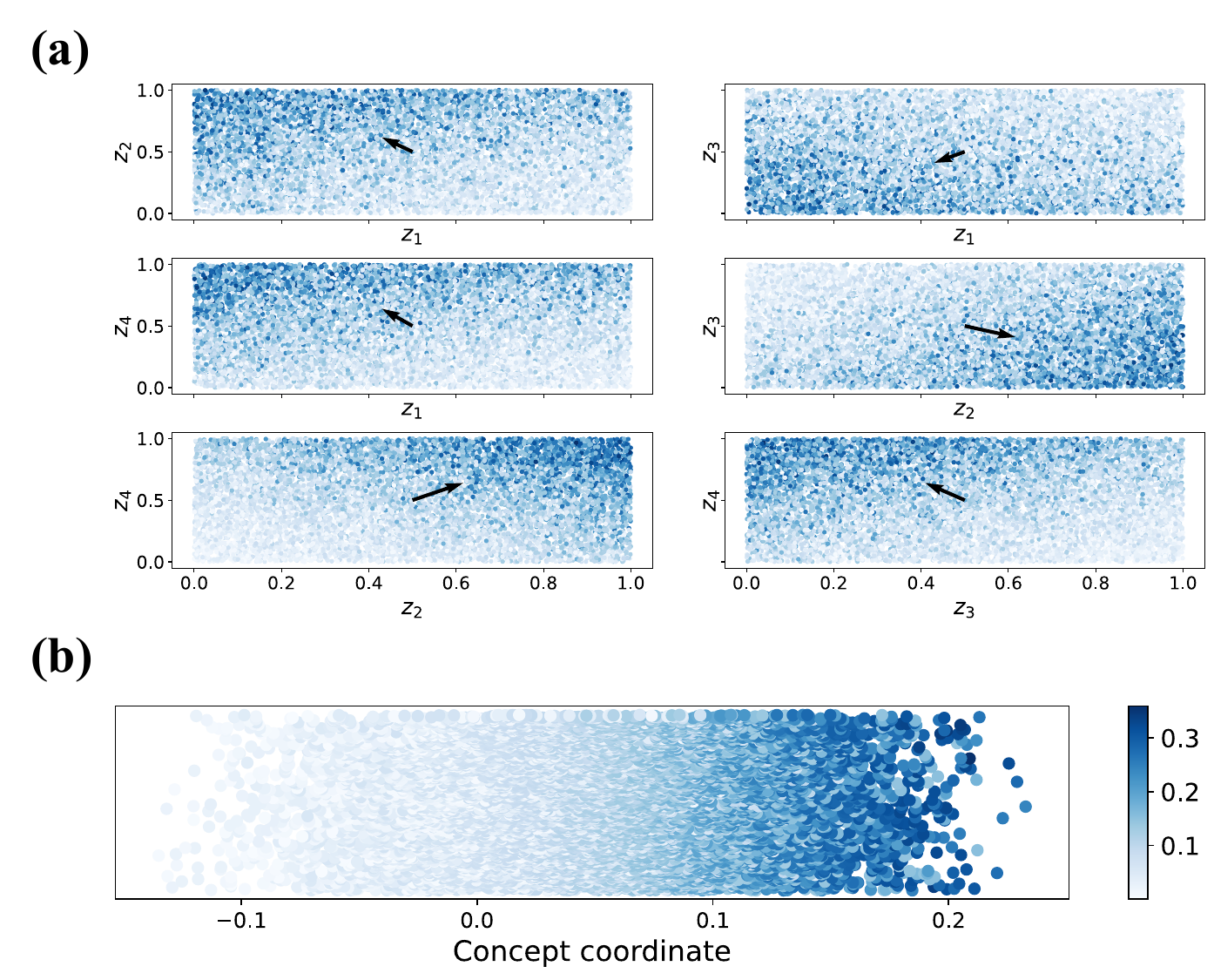}
    \caption{(a) The concept vector visualized in the four dimensional latent space, with $z_1$,$z_2$,$z_3$, and $z_4$ denoting the individual latent space dimensions and the arrow indicating the direction of maximum W variance.
        (b) The projection of the latent codes along the concept vector and colored by the W content in their corresponding generated compositions.
        Vertical axis is random noise to reveal the range of W values at each concept coordinate}
        \label{fig:concept_vector}
\end{figure}

To first verify the similarity between the predicted W content from the LR model and the actual W content in the compositions, we once again relied on the average $R$ value from 10-fold cross-fold validation (CV).
The mean value of $R=0.84$ is a strong indicator that the LR model is indeed able to find a direction in the latent space that captures the change in W content in a better way than any of the individual dimensions of the latent space.
The LR model was then refit using all the 10,000 generated samples, and the coefficient from the model was then finalized as the concept vector.
Fig.~\ref{fig:concept_vector} shows the concept vector obtained from the LR model visualized in the latent space, which clearly shows that the concept vector points towards the direction in the latent space which corresponds to region of higher W content.

\begin{figure}[ht]
    \centering
    \includegraphics[width=0.6\textwidth]{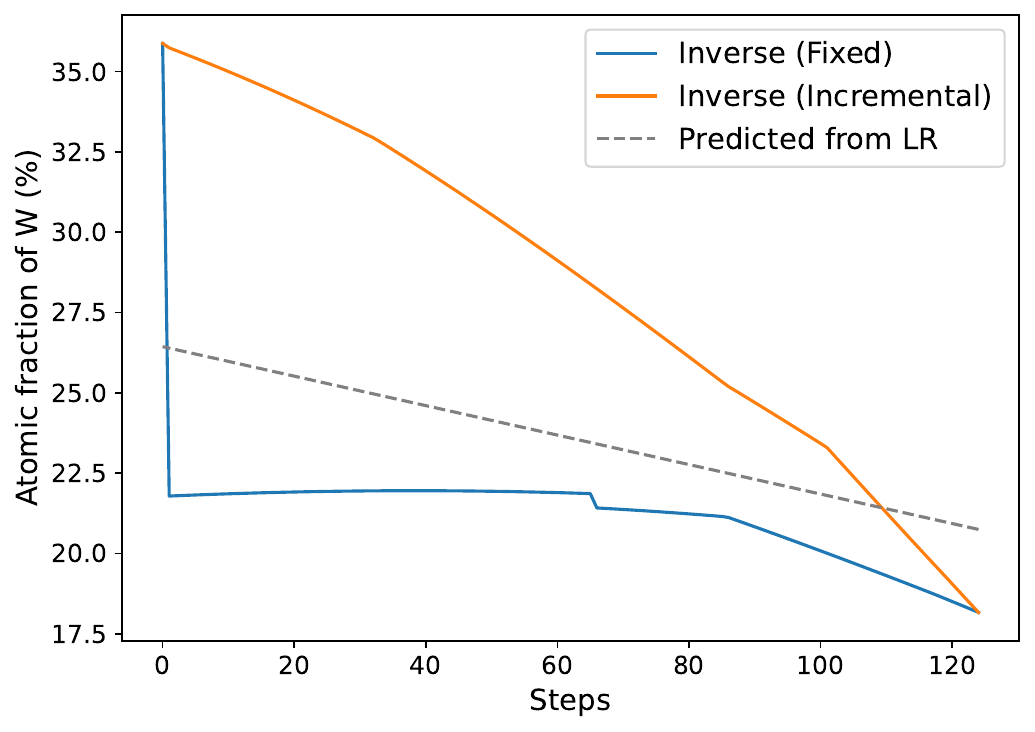}
    \caption{The change in the W content of the generated compositions as we proceed along the concept vector (illustrated in Fig.~\ref{fig:concept_vector}), for the fixed UTS and incremental UTS conditioning cases. The grey dashed line shows the predicted W content for the corresponding latent codes at each step}
    \label{fig:inv_w_content}
\end{figure}

Using the concept vector, we then generated latent codes along the direction of $-v_W$ concept vector since we are interested in generating compositions with lower W content.
Thus, starting with the latent code of the $C_0$, we obtain the latent code of each subsequent composition in the interpolation path using the formula,
\begin{equation}
    z' = z + n \alpha v,
\end{equation}
where $z$ is the latent code of the starting composition, $z'$ is the modified latent code, $\alpha$ is the step size controlling the number of intermediate compositions to be generated (discretizing the continuous $z$ space), and $n$ is the number of steps. 
Here we have chosen $n$ and $\alpha$ to be 125 and 0.01 respectively, though what really matters is their product (i.e., how far from the original composition we move in latent space).
As we are interested in simultaneously increasing the UTS at $1200^{\circ}$C, we also need to set the cGAN condition to a higher value than the UTS of the original composition.
Here we have chosen the target condition to a fairly high value of 0.9 GPa, and there are two ways in which we can implement this condition in practice: 
\begin{enumerate}
    \item \textbf{Fixed condition.} By keeping the UTS fixed for all the intermediate samples, in which case the generator is only concerned with finding the sample with the maximum possible UTS, leading to discontinuous changes in the W content in the generated samples. This is the canonical inverse design process.
    \item \textbf{Incremental condition.} By gradually increasing the UTS values for each subsequent sample by creating equally spaced intervals between the starting and end UTS values equal to the number of intermediate compositions desired. In this case, the generated samples are expected to show a more gradual change in the W content. This construction provides a closer analogue the rational design process.
\end{enumerate}
In both cases, the elements of the generated compositions having atomic fraction less than 3\% generated candidates were set to zero and the composition was normalized to ensure that the atomic fraction sums to unity.

Fig.~\ref{fig:inv_w_content} shows how the atomic fraction of W in the generated compositions change as we proceed along the concept vector for both the fixed UTS and incremental inverse methods. 
As expected, keeping a fixed and sufficiently high UTS value as condition resulted in a discontinuous change in the W content after the first step (from 36 to around 21.5).
Compared to this, the incremental increase in the conditioning UTS value results in a much smoother change in the W atomic fraction, which is in accordance with what we had expected.
While the LR model predicts the atomic fraction of W in the final composition to be around 21, the final W atomic fraction was observed to be 18.
The final composition for both these cases was the same --  \ce{Cr7Hf25Mo14Nb3Ti28W18Zr5}.

\begin{figure}[ht]
    \includegraphics[width=\textwidth]{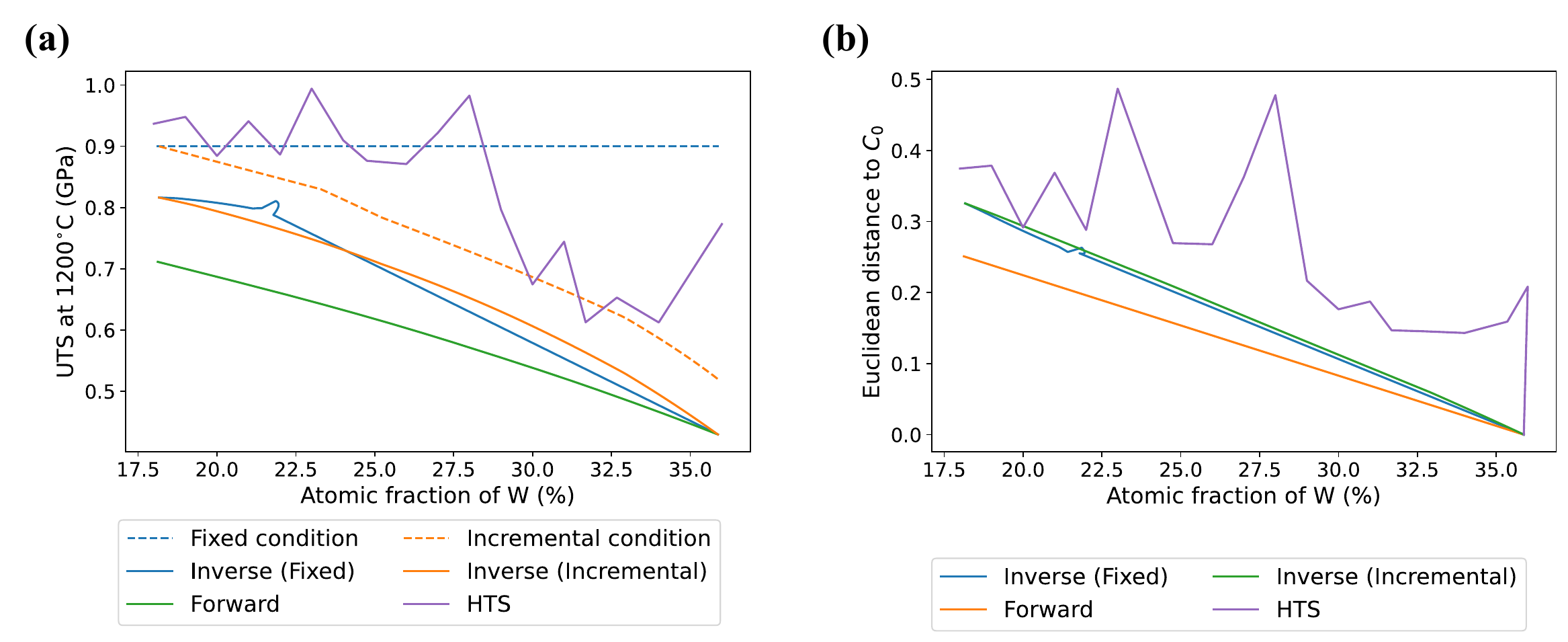}
    \caption{(a) The UTS values at $1200^{\circ}$C for the candidates with reduced W generated from the inverse and forward design methods. The latent codes for the candidates generated by the two inverse design methods were determined by scanning over the concept vector from Fig.~\ref{fig:concept_vector} using the strategy illustrated in Fig.~\ref{fig:inv_w_content}. The blue and orange dashed lines show the conditioning value of UTS provided during the fixed and incremental inverse methods respectively.
    (b) The euclidean distance of the candidates to $C_0$ with the reduction of W atomic fraction}
    \label{fig:uts_compare}
\end{figure}

To provide a direct comparison to the inverse scheme, we devised a forward local search in the design space to reduce the W content in $C_0$ while simultaneously trying to maximize the UTS at $1200^{\circ}$C.
We also used a high-throughput search (HTS) scheme to find a suitable candidate from the training data.
To provide a parallel to the inverse and forward search cases, the HTS was restricted to only those compositions in the training data corresponding to the range of W values used in the previously described methods.
Fig.~\ref{fig:uts_compare}a shows the UTS of the candidates generated by the two inverse designs cases and the forward design methods.
The dashed blue and orange lines show the UTS values provided as conditioning to the cGAN for the fixed and incremental cases respectively, while the solid blue and orange lines are the predicted UTS values of the generated candidates from the UTS surrogate model.
As both the fixed and incremental inverse method results in the same final composition, the UTS values from these two cases are the same. 
However, the fixed UTS case results in not only an abrupt change in W content, as was seen in Fig.~\ref{fig:inv_w_content}, but also results in candidates with higher UTS than those generated by the incremental UTS case.
This is also expected, as the provided condition is always higher in the fixed UTS case compared to the case of incremental UTS, as can be seen by the dashed blue and orange lines in Fig.~\ref{fig:uts_compare}.
Surprisingly, even though the atomic fraction of W is the same, the forward search scheme results in a completely different final composition, \ce{Cr20Hf18Mo24Nb6Ti7W18Zr7}, which also has a lower UTS value at $1200^{\circ}$C than the candidate from the inverse design cases.
The HTS scheme results in a very noisy search over the training dataset, but overall results in candidates with higher UTS values for any given W atomic fraction, with a final composition of \ce{Cr4Hf30Mo29Ti19W18}.
However, it should also be kept in mind that the limit of the UTS values for the inverse design methods was chosen to be 0.9, and higher value of the condition could theoretically have resulted in a higher UTS value of the final candidate from the inverse methods.
Additionally, from Fig.~\ref{fig:uts_compare}b, it can be observed that the change in the euclidean distance of the candidates from the inverse and the forward search cases are smoother than those for the candidates from HTS. 
This clearly indicates that HTS has access to discrete compositions that do not smoothly track with the euclidean distance. 
As a result, the dominant solutions from HTS jump around between high UTS, high distance and low UTS, low distance.
This, combined with the results from Fig.~\ref{fig:uts_compare}a indicates that the higher UTS obtained from the HTS scheme occurs when the scheme can not find candidates with a low euclidean distance to $C_0$, and when it can, it results in a candidate with a low value of UTS.

\begin{figure}[ht]
    \centering
    \includegraphics[width=\textwidth]{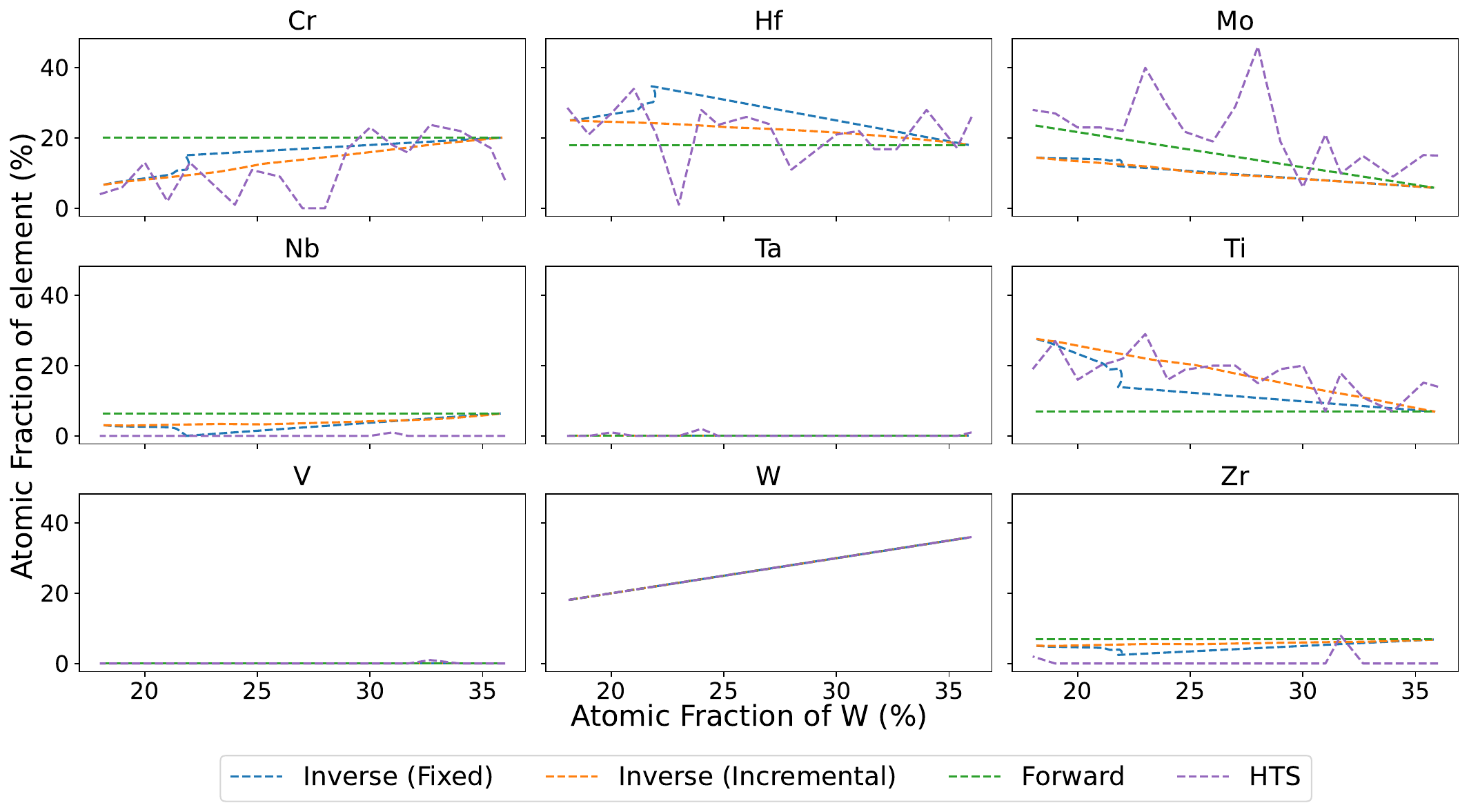}
    \caption{The change in atomic fractions of the different elements using the different design methods. The inverse and the localized forward search methods lead to smoother changes in the atomic fractions, while the HTS method leads to discontinuous changes}
    \label{fig:at_frac}
\end{figure}

Fig.~\ref{fig:at_frac} shows the evolution of the atomic fractions of the elements in generated compositions by the inverse and forward methods.
Note that the nearly superimposed W between methods is enforced by the design process (since W content was a design variable).
For the forward, inverse, and HTS cases, the Ta and V atomic fractions remain unchanged.
The matching near-zero Ta and V content between the design methods indicates that the surrogate models do not favor these elements for achieving the properties of interest.
For both the inverse and HTS cases, the atomic fraction of Cr, Nb, and Zr go down, while those of Hf, Mo, and Ti go up.
However, the forward method results in only an increase in the Mo atomic fraction, with all the other elements remaining unchanged.
The changes in the atomic fraction for the HTS case is especially noisy, which is a consequence of the Pareto optimal candidates not varying smoothly along the jagged line shown in Fig.~\ref{fig:uts_compare}. 
We believe this is the result of discretization in the search space (i.e., the synthetic data set); the sharp transitions between compositions indicate that the HTS scheme does not perform optimization and is subject to depletion in certain regions of the space.
This also indicates that the results of HTS will strongly depend on the scheme used to construct the search space.

\subsection{Multiple condition design case}

In addition to UTS, we also desire that the designed alloys meet certain additional criteria.
Particularly, in order for the designed alloy to be practical, it is preferable for it to have densities ranging between 8 and 10 g/$\textrm{cm}^3$ and be as affordable as possible.
For this scenario, the cGAN architecture was modified to include price and density of the refractory HEAs as conditions, resulting in the input to the cGAN model now being a seven-dimensional vector (four latent dimensions plus three conditioning value).
All the remaining training aspects of the cGAN were identical to the single condition case.
Similar to the single condition case, the cGAN results were compared against HTS. 
As the forward search scheme was the worst performing for the single condition case and required the most manual effort to configure, we opted to not consider it in the multi-condition study and instead decided to use a Multi-Objective Optimization (MOO) scheme for comparison. 

\begin{table}[ht]
\centering
\caption{The generated candidates from the different forward and inverse schemes. For the inverse candidates, the number inside the parenthesis indicate the conditioning value used for generation}
\begin{tabularx}{\linewidth}{|c|X|X|X|X|}
\toprule
\textbf{Composition} & \textbf{Method} & \textbf{Predicted UTS (GPa)} & \textbf{Predicted Price (\$/kg)} & \textbf{Predicted Density (g/$\textrm{cm}^3$)}
\\
\midrule
\ce{Mo50Nb5Ti37W5Zr3} & HTS & 1.12 & 31.54 & 8.24 \\ 
\ce{Cr11Mo52Nb20Ti14Zr3} & MOO & 1.0  & 41.76 & 8.53 \\ 
\ce{Cr25Mo49Nb4Ti15W7} & cGAN & 1.04 (1.2) & 29.65 (30) & 9.09 (9) \\
\ce{Cr22Mo58Ti16W4} & cGAN & 1.13 (1.4) & 28.61 (30) & 8.92 (9)\\
\ce{Cr19Mo54Ti27} & cGAN & 1.14 (1.4) & 26.60 (31.54) & 8.05 (8.24)\\
\bottomrule

\end{tabularx}
\label{tab:multi_obj}
\end{table}

Table~\ref{tab:multi_obj} shows the candidates generated from the different schemes considered for the design of refractory HEAs satisfying multiple conditions. 
Note that while all the design schemes are applied to the 9 refractory elements described above, the different methods result in different subsets of those elements.
The candidate from MOO has the lowest UTS value out of all the generated compositions, while also having the highest price, indicating poor convergence.
The HTS candidate from the training set is able to match the provided conditions fairly well, but finding it required some manual tuning, as was described in Section~\ref{hts_meth}.
The cGAN generator does not require any additional manual tuning post training and only needs to be provided with the target properties to generate the candidates. 
For the cGAN compositions, a higher UTS value needs to be provided as conditioning as the cGAN generator tends to underfit at the extreme ends of the conditioning values (in agreement with \cite{Debnath2021}).
For all the three cGAN compositions, we can see that the predicted and the provided conditioning values are very similar to the condition provided and indicates the generator's power to satisfy multiple criteria.

Generation of new candidates with different targets is also very straightforward and only involves changing the relevant conditioning values.
This can be seen for the candidates \ce{Cr22Mo58Ti16W4} and \ce{Cr19Mo54Ti27}, which were obtained by changing only the UTS and all three conditions respectively.
These two candidates are also especially interesting, as they are able to show higher UTS and lower price compared to the best candidate from the forward methods (HTS and MOO).
Moreover, changing the target property values do not affect the inference time of the generator, with the generator taking only microseconds ($76.4 \pm 1.5 \mathrm{\upmu s}$) to generate a composition.
This is a major advantage over the forward models, which can take seconds (HTS) to minutes (MOO) to converge on suitable candidates for different targets.
Additionally, the search space and objectives need careful consideration in order to obtain meaningful results for these forward methods (e.g., constraints in MOO or the reference point for hypervolume calculation in HTS).

\section{Conclusion}

In this study, we have directly compared forward (search-based) and inverse (generative-modeling-based) design schemes as applied to single and multi objective refractory HEA design.
For the single condition case, we have proposed a way to control the elemental fraction in candidate refractory HEAs generated by a cGAN model using latent space arithmetic while maximizing the UTS of the generated compositions using conditioning.
Our proposed method allows us to keep the atomic fraction of the elements separate from the conditioning provided to the cGAN, which is necessary to avoid circular logic as the compositions are themselves the cGAN output.
We also compared the inverse design method to a localized forward search scheme and observed that the inverse design method was able to identify a composition with the same constraints but with higher predicted UTS.
While the HTS scheme was able to identify compositions with higher UTS values compared to the cGAN, these often come at the cost of deviating further away from the starting composition.

For the multi condition case, we observed that the candidates from cGAN are able to satisfy the imposed conditions fairly well. 
While the forward methods could find candidates with desired UTS and density, these candidates could not closely match the target price.
Interestingly, when provided with target price and density corresponding to the best candidate from HTS, the cGAN was able to generate a composition with lower price and density than the HTS candidate while showing slightly higher UTS value.
Additionally, the generation of new candidates with different targets is instantaneous with the trained cGAN generator, but forward methods like HTS and MOO need to be performed from scratch each time.
Moreover, the overhead for the forward search and HTS method increases when considering more material properties or imposing additional constraints. 
For the inverse design case, addition of new conditions is fairly straightforward and generation of new candidates is fast (i.e., inference with a small neural network).

Therefore, these results indicate that inverse methods like cGAN offer certain benefits over the forward design methods outlined in the study.
However, there are potential scenarios that may require attention.
For instance, while we have only considered the case where the desired concept vector governing the change in atomic fraction of W could be identified, there may be scenarios where the desired concept vector cannot be readily found in the latent space.
Such a case study (and an evaluation of the prevalence of such cases) could be the subject of future research.
Additionally, another interesting and relevant direction of research would be to benchmark the inverse method against more sophisticated forward search methods like Genetic Algorithm and Bayesian Optimization \cite{whitley1994genetic, frazier2018tutorial}.
Ultimately, we believe that the results presented here highlight unique advantages of generative-modeling-based inverse design for materials that should spark additional interest in systematic benchmarking in refractory HEAs and other materials design problems.

\section{Methods}

\subsection{Generative models and latent space}\label{gen_model}

Out of the generative models discussed in Section~\ref{into}, the latent-variable models like VAE and GAN have gained popularity in the recent years due to their ability to reduce a complex, high-dimensional design space into a lower dimensional space (called the latent space) that captures the data distribution governing the original space. 
The idea of the latent space is that each point in this reduced space can be provided to the generative model to produce a novel sample.
The latent space is smooth and continuous by construction (done by choosing a random normal distribution), which allows similar observations to be grouped closer in the latent space than in the original space.

Compared to VAEs, GANs have been found to produce less noisy samples as the VAE training procedure inherently involves injecting noise and relies on a reconstruction loss computed using a metric like mean squared error, which leads to imperfect element-wise measures \cite{bao2017cvae}.
The GAN architecture consists of two deep neural networks (DNNs) - a generator ($G$) that tries to produce realistic looking samples from random noise drawn from a normally distributed latent space and a discriminator ($D$) that tries to identify if the sample was drawn from the training domain or from the generator.
The two models are trained in tandem, with the generator creating batches of fake samples, which are then passed on to the discriminator along with batches of real sample to be classified as real or fake.
The generator is updated based on how well the generated samples were able to fool the discriminator. Simultaneously, the discriminator is also updated to become better at correctly identifying real/fake samples. 
Therefore, both these networks are competing against each other  during training, and we eventually arrive at a state where the generator is able to produce samples indistinguishable from the real data and the discriminator has an near equal probability of classifying a sample as real or fake.
The discriminator can then be discarded and the trained generator can then be used to generate realistic fakes.

\begin{figure}[ht]
     \centering
        \includegraphics[width=0.5\textwidth]{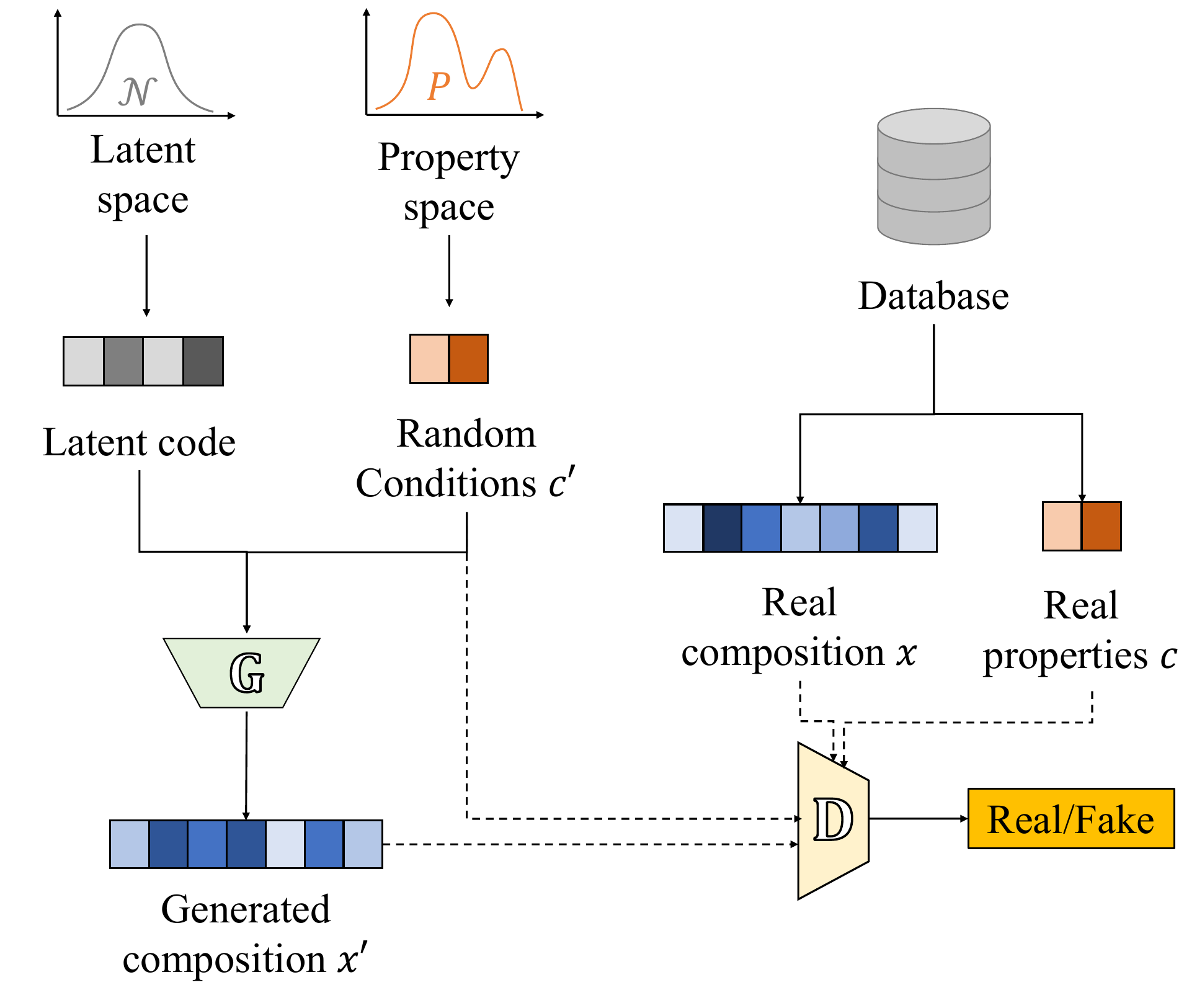}
        \caption{The conditional GAN architecture is a variation of the GAN architecture, with the modification that an additional conditioning vector is provided as an input to both $G$ and $D$}
        \label{fig:gan_to_cgan}
\end{figure}

However, there is no mechanism to control the output of a GAN, and as a consequence, many candidates need to be generated before we obtain one with desirable properties.
This can be rectified by modifying the architecture slightly to convert a GAN to a conditional GAN (cGAN), a variant of GAN that uses an additional auxiliary information (also referred to as conditioning vector) as input to both the generator and the discriminator, as can be seen in Fig.~\ref{fig:gan_to_cgan}.
The conditioning vector is usually chosen to be a trait of the generated sample that we wish to control - for images, these can be the image class, while for the problem of alloy design, these can include continuous-valued material properties like hardness and fracture toughness \cite{Debnath2021}. 
This allows the generator to learn a multi-modal mapping from the input to the output domain by using the context provided by the conditioning vector \cite{aggarwal2019regression}.

In this work, we have utilized a cGAN model with 4 fully connected layers (3 hidden layers + 1 output layer), conditioned on the properties of interest.
The property values were standardized and subsequently sampled using its empirical probability distribution. 
We have used a vector array with length equal to the number of unique elements in our training dataset of the to represent the alloy composition.
Since the order of arrangements of the elements do not matter when using fully connected layers, we have chosen the alphabetical order to arrange the elements in the representation.
Each component of the vector corresponds to the atomic fraction of a particular element.
This particular representation was chosen as we recently found that using more sophisticated representation schemes like the Periodic Table Representation does not lead to drastically improved performance when used for deep learning tasks \cite{debnath2022representations}.
The latent dimension of the cGAN is set to four, with the number of properties being used as condition determining the dimension of the input vector to the generator.
The generator outputs a vector, whose components correspond to the atomic fraction of elements used in training set.
The learning rate for both the generator and the discriminator were set to $10^{-3}$.
The model was implemented using the \texttt{pytorch} Python package \cite{pytorch2019}.

\subsection{Training data} \label{training}

Generally, most traditional deep learning models need a sizeable dataset for training (around $10^4 - 10^5$ observations).
However, such a large dataset might be difficult to construct, especially in materials science, where it is usually expensive and and time consuming to generate a large volume of data.
Therefore, these smaller datasets (around $10^2 - 10^3$ observations) are not ideal for being deployed directly for deep learning tasks.
These datasets have incomplete property sets due to the expense and complexity of performing experimental characterization. 
For mechanical properties some tests are also destructive which make it difficult or impossible to obtain complete property sets for a single sample. 
A way to circumvent this issue is by using a larger synthetic dataset constructed by combining design of experiments (DOE) and phenomenological models or ML surrogate models trained on the smaller dataset. 
The synthetic dataset thus allows us to test the design methods irrespective of the data sparsity challenge coming from experimental characterization.
An added benefit of using surrogate models is that they can act as an intermediate step before experimental validation and can be used as a quicker and cheaper way to compare the performance of the generative model.

In this work, we have created a synthetic dataset of 56,837 observations of compositions made from different combination of 9 elements (Cr, Hf, Mo, Nb, Ta, Ti, V, W, Zr) obtained from DOE with three properties (UTS at $1200^{\circ}$C (GPa), price (\$/kg) and density (g/$\textrm{cm}^3$).
The dataset contained compositions with number of elements ranging from 2 to 6, with the median being 5.
The density and price of the refractory HEAs were determined using phenomenological models, where each of these properties of the alloy were calculated using a linear combination of the corresponding pure elemental properties from DFT calculations \cite{Chong_2021}.
The UTS at $1200^{\circ}$C was obtained from a ML surrogate model trained on the UTS values of HEAs, taken from the ULTERA database \cite{ULTERA_ZenodoDataset}. 
The training dataset has 268 observations using 18 elements of the periodic table. 
The surrogate model was chosen to be a shallow neural network with one hidden layer with 50 neurons and the ELU activation function.
The input to the model is a vector containing information like the atomic fractions of the elements present in the composition, processing parameter (a one-hot vector 
of length 1 to signify if the measurement was done on the as-cast alloy or after putting it through some heat treatment), phase (a one-hot vector of length 3 to signify if the phase of the alloy is a pure solid solution, mixture of solid solutions, or intermetallics) and the temperature at which the UTS was measured.
Both the input and the UTS values were standardized before training the model using the \texttt{StandardScaler} function from \texttt{sklearn} library \cite{sklearn}.
While typically the accuracy of the surrogate model would be of critical importance in a material design study, here we treat the surrogate model as the ground truth for lack of ready access to validation experiments.
That is, we take a comparable prediction by the surrogate model prediction as validation of the various design schemes evaluated here (with the intention to benchmark the design schemes themselves), and leave experimental validation to future work (with the intention to use the schemes to design new materials).

\subsection{Forward search method}

The forward local search method was performed in the design space to reduce the W content in $C_0$ while simultaneously trying to maximize the UTS at $1200^{\circ}$C.
We iteratively remove a fixed amount of W from the composition, starting with $C_0$, and redistribute it among the other eight elements considered to generate the intermediate candidates.
At each iteration, the candidates are first evaluated to ensure that the atomic fractions sum to unity, after which the UTS at $1200^{\circ}$C is calculated using the surrogate model.
The candidate with the highest UTS is then set at the starting composition for the next round of evaluation.
This iterative process continues until we have reached a composition with the desired atomic fraction of W.
The described search scheme allows us to stay close to $C_0$ instead of sampling the entire design space for the composition that corresponds to the global optimum.
For this work, we have used 0.05 as the value by which the W content is reduced, and the final W content to be achieved was set as the W content of the final composition obtained by the inverse process.

\begin{figure}[ht]
    \centering
    \includegraphics[width=\textwidth]{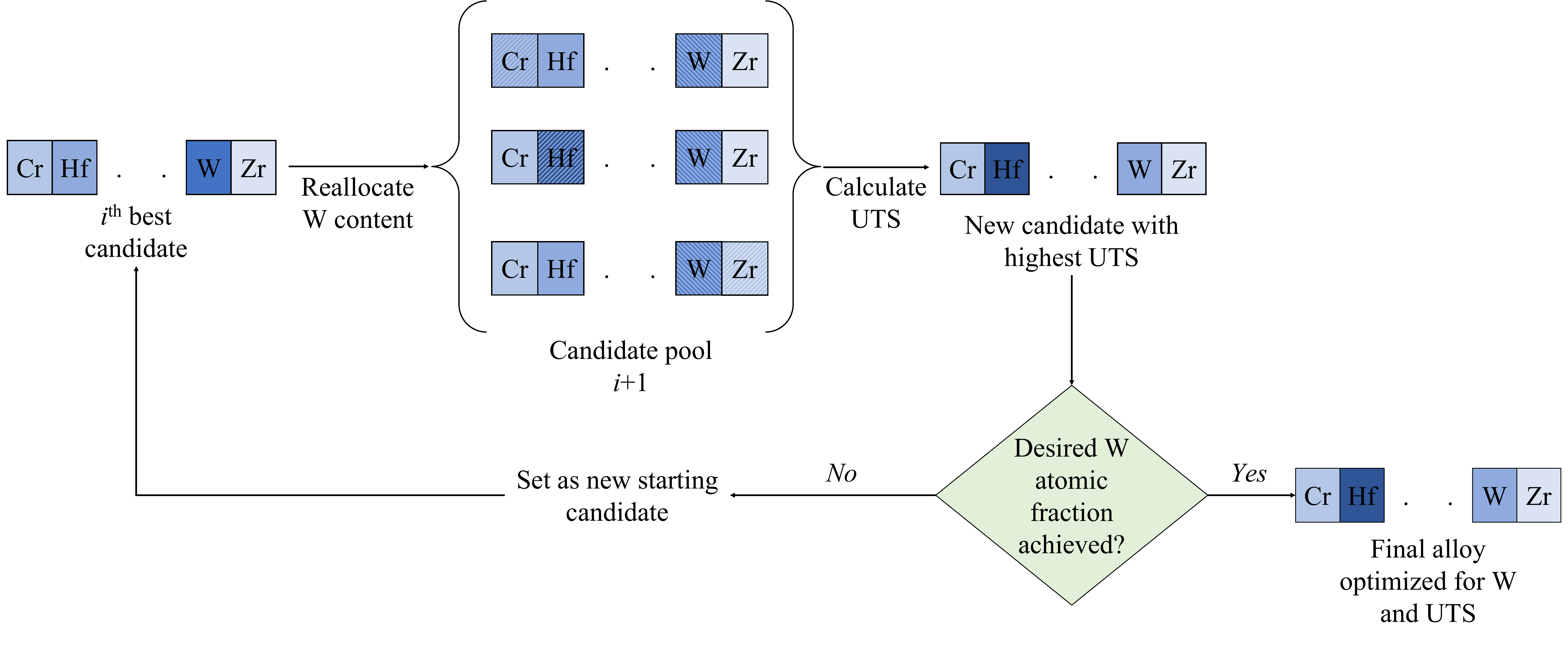}
    \caption{A schematic of the forward search method. The intensity of blue indicates the atomic fraction of the corresponding elements. At each step, an ensemble of candidates are generated by removing a certain amount of W from the starting composition and adding it to each of the elements considered. The candidate with the highest UTS becomes the starting composition for the next step, until the desired W atomic fraction is achieved}
    \label{fig:forward_search_schematic}
\end{figure}

\subsection{High-throughput search (HTS) method} \label{hts_meth}

For the single condition case, to provide a parallel to the inverse and forward search cases, the HTS was restricted to only those compositions in the training data corresponding to the range of W values used in the previously described methods. 
The first step involved finding the non-dominated solutions that lie at the Pareto front of the W-UTS property space (minimizing W while maximizing UTS). 
To ensure that the training set candidates did not deviate greatly from $C_0$, we included the Euclidean distance of the training set compositions to $C_0$ as a third dimension for calculating the Pareto frontier.
The hypervolume calculations were then performed on the Pareto efficient solutions using the \texttt{pymoo} Python package \cite{pymoo}.
The reference point chosen for the hypervolume calculations were placed at: the W content at each step, -UTS of $C_0$, and Euclidean distance of 1.
The candidate from the training set with the highest hypervolume in each bin of W atomic fractions was then selected as the optimal.

For the multi-condition case, no restriction was placed on the search space as we aren't concerned with controlling the W content in the composition in this case.
The non-dominated solutions were calculated in a three dimensional space defined by (-UTS, density - 9~$\mathrm{g/cm^3}$, price).
The value of 9~$\mathrm{g/cm^3}$ was chosen based on external factors for our alloy design task. 
Initial attempts in conducting the search in this space resulted in selection of binary alloys as candidates.
However, as we are in search of a multi-component alloy, we used 4 - the number of elements ($N_\mathrm{ele}$) of the alloy as an additional dimension for the Pareto efficient solution calculations.
The hypervolume calculations were then performed using the reference point of (-1~$\mathrm{MPa}$ for UTS, 10~$\mathrm{g/cm^3}$ for density, 30~$\$/kg$ for price, and 4 for $N_\mathrm{ele}$.
The alloy with the highest hypervolume was then chosen as the candidate.

\subsection{Multi-objective optimization (MOO)}

The multi objective optimization was performed using the \texttt{pymoo} \cite{pymoo} Python package.
The objectives of the optimization scheme were chosen to be to simultaneously maximize the UTS at $1200^{\circ}$C and minimize the price.
Additionally, constraints  were also provided such as atomic fractions should sum to unity and density should be in the range 8-10~$\mathrm{g/cm^3}$.
The RNSGA2 algorithm was used with a reference point of (-1~$\mathrm{MPa}$ for UTS, 30~$\mathrm{\$/kg}$ for price).
The algorithm was initialized with a population of 500 sampled using Latin Hypercube method, with each candidate generating 50 offspring in the next generation using crossover and mutation.
The algorithm was terminated after 200 generations.

\section*{Declarations}

\subsection*{Acknowledgements}

The present work is based upon work supported by the Department of Energy / Advanced Research Projects Agency – Energy (ARPA-E) under award No DE-AR0001435. 
The authors would also like to thank John Shimanek, Christopher DeSalle, and Douglas Wolfe for helpful discussions.

\subsection*{Conflict of Interest}
The authors declare that they have no known competing financial interests or personal relationships that could have appeared to influence the work reported in this paper.

\subsection*{Software and data availability}
A static snapshot of the code and data used to generate the results in this work are publicly available on \href{https://doi.org/10.5281/zenodo.8061193}{Zenodo}~\cite{Debnath2023software}.

\bibliography{bibliography}

\end{document}